\documentclass[aps,prb,twocolumn,showpacs,amsmath,amssymb,superscriptaddress]{revtex4}
\usepackage{graphicx}
\usepackage{amssymb}
\usepackage{natbib}
\usepackage{color}

\newcommand{\beq}{\begin{eqnarray}}
\newcommand{\eeq}{\end{eqnarray}}

\begin{document}
\title{Unusual suppression of a spin resonance mode with magnetic field in underdoped NaFe$_{1-x}$Co$_x$As: Evidence for orbital-selective pairing}
\author{Yu Song}
\email{yusong@berkeley.edu}
\affiliation{Department of Physics and Astronomy, Rice University, Houston, Texas 77005, USA}
\affiliation{Department of Physics, University of California, Berkeley, California 94720, USA}
\author{Guotai Tan}
\affiliation{Center for Advanced Quantum Studies and Department of Physics, Beijing Normal University, Beijing 100875, China}
\author{Chenglin Zhang}
\affiliation{Department of Physics and Astronomy, Rice University, Houston, Texas 77005, USA}
\author{Rasmus Toft-Petersen}
\affiliation{Helmholtz-Zentrum Berlin f\"{u}r Materialen und Energie, Hahn-Meitner-Platz 1, D-14109 Berlin, Germany}
\author{Rong Yu}
\affiliation{Department of Physics and Beijing Key Laboratory of Opto-electronic Functional Materials and Micro-nano Devices, Renmin University of China, Beijing 100872, China}
\author{Pengcheng Dai}
\email{pdai@rice.edu}
\affiliation{Department of Physics and Astronomy, Rice University, Houston, Texas 77005, USA}
\affiliation{Center for Advanced Quantum Studies and Department of Physics, Beijing Normal University, Beijing 100875, China}

\begin{abstract}
We use inelastic neutron scattering to study the fate of the two spin resonance modes in underdoped superconducting NaFe$_{1-x}$Co$_x$As ($x=0.0175$) under applied magnetic fields. While an applied in-plane magnetic field of $B=12$ T only modestly suppresses superconductivity and enhances static antiferromagnetic order, the two spin resonance modes display disparate responses. The spin resonance mode at higher energy is mildly suppressed, consistent with the field effect in other unconventional superconductors. The spin resonance mode at lower energy, on the other hand, is almost completely suppressed. Such dramatically different responses to applied magnetic field indicate distinct origins of the two spin resonance modes, resulting from the strongly orbital-selective nature of spin excitations and Cooper-pairing in iron-based superconductors.
\end{abstract}


\maketitle


\section{Introduction}

Iron-based superconductivity appears in proximity to antiferromagnetic (AF) order \cite{DScalapino2012_RMP,PDai2015_RMP,Inosov}, with spin fluctuations central to its pairing mechanism \cite{DScalapino2012_RMP}. Like other families of unconventional superconductors \cite{JRossat-Mignod1991_PC,CStock2008_PRL}, an intense spin resonance mode (SRM) is observed by inelastic neutron scattering in the superconducting state of iron-based superconductors \cite{ADChristianson2008_Nature}, indicative of sign-reversed superconducting order parameters on different parts of the Fermi surface \cite{hirschfeld}. The SRM is believed to be an electron-hole spin-triplet bound state inside the superconducting gap \cite{MEschrig2006,YSidis2007}, and its intensity acts as a proxy for superconducting pairing correlations \cite{Dai2000}. Under in-plane magnetic fields well below the upper critical field, intensity of the SRM is observed to be only mildly suppressed \cite{PBourges1997,Dai2000,JSWen2010_PRB,JZhao2010_PRB,SLi2011_PRB}, consistent with the notion that intensity of the SRM tracks the superconducting order parameter.

The electronic structure of iron-based superconductors is dominated by Fe 3$d$ $t_{2g}$ orbitals near the Fermi level, with hole-like Fermi surfaces at the zone center $\Gamma$ and electron-like Fermi surfaces at the zone corner $M$, and the superconducting order parameter changes sign between these quasi-nested Fermi surfaces \cite{hirschfeld,GRStewart}. The presence of multiple Fe 3$d$ orbitals near the Fermi level adds an orbital degree of freedom to the physics of iron-based superconductors, resulting in varying orbital characters on different parts of the Fermi surfaces [Fig.~1(a)] \cite{MYi_npj_QM}, and orbital-dependent strengths of electronic correlations \cite{ZPYin_NM,deMedici_PRL}. Such strong orbital-dependence then leads to orbital-selective Mott phases \cite{MYi_PRL2013,RYu_PRL2013} and orbital-selective Cooper pairing \cite{EMNica_npj_QM,POSprau} in iron-based superconductors.

The orbital degree of freedom also manifests in spin excitations, as exemplified by orbital-selective spin excitations in LiFe$_{1-x}$Co$_x$As \cite{YLi_PRL2016} and double SRMs observed in underdoped NaFe$_{1-x}$Co$_x$As \cite{CZhang2013_PRL}. The double SRMs in underdoped NaFe$_{1-x}$Co$_x$As is suggested to result from orbital-dependent pairing, with superconducting gaps along the electron-like Fermi surface associated with different orbitals exhibiting differing superconducting gaps [Fig.~1(b)], and thus SRMs associated with different orbitals also appear at different energies.
Double SRMs with different spin space anisotropy
are also observed in optimally electron- \cite{PSteffens2013_PRL}, hole- \cite{CZhang2013_PRB}, and isovalent-doped BaFe$_2$As$_2$ \cite{DHu2017_PRB}, although the two SRMs in these materials are not well-separated in energy, and are only revealed through neutron polarization analysis.
In addition to orbital-selective pairing, the two modes have also been suggested to arise from the presence of static or slowly fluctuating magnetic order \cite{WLv2014_PRB,MWang2016_PRB} or due to spin-orbit coupling (SOC) \cite{SVBorisenko_NP2016} that lifts spin-space degeneracy of the SRM \cite{MMKorshunov2013_JSNM,YSong2016_PRB,DDScherer2017}.  Underdoped superconducting NaFe$_{1-x}$Co$_x$As offers a unique opportunity to probe the nature of its double SRMs using an applied magnetic field for several reasons. First, it exhibits competing superconductivity and AF order that can be tuned by a field accessible in a neutron scattering experiment. Second, the double SRMs are well-separated in energy and can be resolved without polarization analysis.
Finally, angle-resolved photoemission spectroscopy (ARPES) measurements revealed nodeless but anisotropic superconducting gaps at zero-field \cite{QQGe_PRX2013}, which may arise from orbital-selective pairing.

In this work, we present an inelastic neutron scattering study of magnetic order and excitations in underdoped NaFe$_{1-x}$Co$_x$As ($x=0.0175$) \cite{CZhang2013_PRL} under an in-plane magnetic field. We find with a field of $B=12$ T, superconductivity is modestly suppressed while AF order becomes slightly enhanced. Of the two SRMs in underdoped NaFe$_{1-x}$Co$_x$As \cite{CZhang2013_PRL}, the mode at higher energy is modestly suppressed, in line with the similar modest suppression of superconductivity; the SRM at lower energy, however, is strongly suppressed and becomes indiscernible for $B\gtrsim10$ T. The complete suppression of a SRM under magnetic field while superconductivity persists is highly unusual, and could result from strongly orbital-selective pairing in iron pnictide superconductors.
Our observations suggest superconducting gaps on orbitals with weak pairing strengths can be suppressed by magnetic fields well below the upper critical field, while bulk superconductivity persists due to orbitals that exhibit stronger pairing. Our work provides strong evidence for orbital-selective pairing and spin excitations in iron-based superconductors.

\begin{figure}[t]
	\includegraphics[scale=.45]{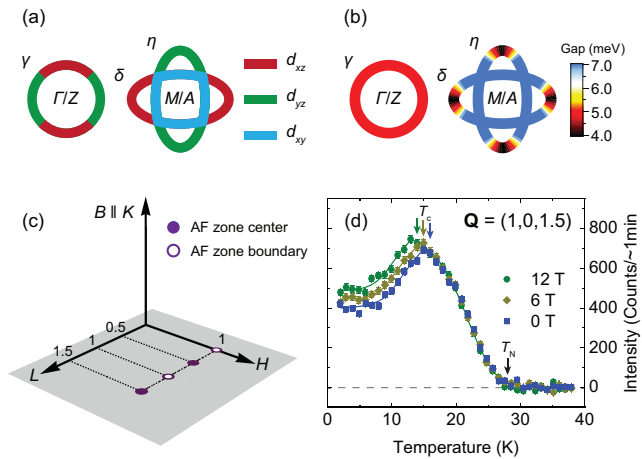}
	\caption{
		(Color online) (a) Schematic Fermi surface of underdoped NaFe$_{1-x}$Co$_x$As, with dominant orbital contributions marked by different colors, based on ARPES measurements \cite{YZhang_PRB2012,MYi_NJP2012}. (b) Schematic of momentum-dependent superconducting gaps in underdoped NaFe$_{1-x}$Co$_x$As, adapted from previous work~\cite{QQGe_PRX2013}. The in-plane zone center is $\Gamma$ or $Z$ depending on $k_z$, and the in-plane zone corner is $M$ or $A$ depending on $k_z$. (c) Schematic of the $[H,0,L]$ scattering plane. The magnetic field $B$ is along $K$, perpendicular to the scattering plane. (d) Magnetic field dependence of the magnetic order parameter measured at ${\bf Q}=(1,0,0.5)$. A constant background has been subtracted. The solid lines are guides-to-the-eye.
	}
\end{figure}

\section{Results}

Single crystals of NaFe$_{1-x}$Co$_x$As ($x=0.0175$) were grown using the self-flux method and have been previously studied using transport \cite{GTan2016_PRB}, ARPES \cite{QQGe_PRX2013}, and neutron scattering measurements \cite{SVCarr2016_PRB,CZhang2016_PRB,GTan2016_PRB} at zero-field. Inelastic neutron scattering experiments were carried out using the FLEXX three-axis spectrometer at Helmholtz-Zentrum Berlin, Germany. Fixed $k_{\rm f}=1.55$ \AA$^{-1}$ was used for all the measurements, and higher-order neutrons are eliminated by using a velocity selector before the monochromator and a Be filter after the sample. We denote momentum transfer ${\bf Q}=(Q_x,Q_y,Q_z)$ in reduced lattice unit (r.l.u.) as ${\bf Q}=(H,K,L)$, with $H=\frac{aQ_x}{2\pi}$, $K=\frac{bQ_y}{2\pi}$, and $L=\frac{cQ_z}{2\pi}$, using $a\approx b\approx5.57$ {\AA} and $c\approx6.97$ {\AA} appropriate for the orthorhombic magnetically ordered phase of NaFe$_{1-x}$Co$_x$As \cite{SLi2009_PRB,GTan2016_PRB}. In this notation, magnetic Bragg peaks appear at ${\bf Q}=(1,0,L)$ with $L=0.5,1.5,2.5\ldots$, whereas integer $L$-values corresponds AF zone boundaries along $c$ axis [Fig.~1(c)]. Our samples were aligned in the $[H,0,L]$ scattering plane and placed inside a magnet with the field direction perpendicular to the scattering plane along $K$ [Fig.~1(c)].

NaFe$_{1-x}$Co$_x$As ($x=0.0175$) exhibits competing superconductivity and AF order with an ordered moment  $\sim0.03\mu_{\rm B}$/Fe \cite{GTan2016_PRB}. Magnetic field dependence of the AF order parameter is shown in Fig.~1(d) for $B=0$, 6 and 12 T. AF order onsets below $T_{\rm N}\approx28$ K regardless of field, and the intensity above $T_{\rm c}$ is unaffected by applied field. This indicates that unlike in-plane uniaxial pressure \cite{CDhital2012_PRL,YSong2013_PRB,DTam2017_PRB,WWang_PRB2017}, for $T>T_{\rm c}$ applied magnetic field up to $B=12$ T affects neither the AF ordered moment size nor population of the AF domains that order at ${\bf Q}_1=(1,0)$ or ${\bf Q}_2=(0,1)$ in underdoped NaFe$_{1-x}$Co$_x$As. The AF order parameters are reduced with the onset of superconductivity below $T_{\rm c}$, indicative of the competing nature of the two orders in iron pnictides \cite{ADChristianson2009_PRL,JLarsen2015_PRB}. With applied field, $T_{\rm c}$ is reduced from its zero-field value $T_{\rm c}\approx 16$ K to $T_{\rm c}\approx14$ K for $B=12$ T. The modest suppression of $T_{\rm c}$ is consistent with the large upper critical field $B_{\rm c2}\gtrsim40$ T in underdoped NaFe$_{1-x}$Co$_x$As \cite{SGhannadzadeh2014_PRB}. Due to the suppression of superconductivity under applied field, the AF order parameter inside the superconducting state becomes enhanced with applied field. At $T=2$ K, the magnetic intensity becomes $\sim20\%$ stronger for $B=12$ T compared to $B=0$ T. Overall, the effects of a $B=12$ T in-plane  magnetic field on NaFe$_{1-x}$Co$_x$As ($x=0.0175$) appear modest, it reduces $T_{\rm c}$ by $\sim10\%$ while enhancing the AF ordered moment also by $\sim10\%$.

Magnetic field dependence of spin fluctuations at ${\bf Q}=(1,0,0.5)$ is shown in background-subtracted constant-${\bf Q}$ scans in Fig.~2(a) (see Appendix for details on background subtraction). We find the normal state response above $T_{\rm c}$ to be field-independent, similar to other iron-based superconductors \cite{MWang2011_PRB}, and therefore the normal state data collected at different fields are combined (see Appendix for field-dependence of normal state excitations). Below $T_{\rm c}$ at $B=0$ T, we observe SRMs centered at $E_{\rm r1}\approx3.25$ meV and $E_{\rm r2}\approx6.5$ meV, with a valley at $E\approx4.5$ meV that display little or no enhancement below $T_{\rm c}$, in agreement with previous work \cite{CZhang2013_PRL}. Surprisingly, at $B=12$ T the mode at $E_{\rm r1}$ becomes strongly suppressed, while the mode at $E_{\rm r2}$ meV is only mildly suppressed, resulting in a single discernible SRM at $B=12$ T.

\begin{figure}[t]
	\includegraphics[scale=.45]{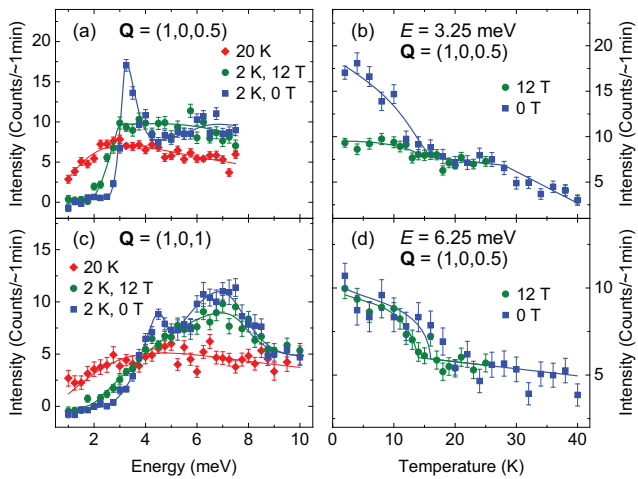}
	\caption{
		(Color online) (a) Background-subtracted constant-${\bf Q}$ scans at ${\bf Q}=(1,0,0.5)$. (b) Temperature dependence at ${\bf Q}=(1,0,0.5)$ and $E=3.25$ meV under $B=0$ T and 12 T, with background subtracted. (c) Background-subtracted constant-${\bf Q}$ scans at ${\bf Q}=(1,0,1)$. (d) Temperature dependence at ${\bf Q}=(1,0,0.5)$ and $E=6.25$ meV under $B=0$ T and 12 T, with background subtracted. The solid lines are guides-to-the-eye. See Appendix for details on background subtraction.
	}
\end{figure}

To verify the dramatically different fate of the SRMs under $B=12$ T, temperature dependence of the two modes are compared at $B=0$ T and $B=12$ T in Figs. 2(b) and 2(d). While both SRMs at $E_{\rm r1}$ and $E_{\rm r2}$ display clear anomalies at $T_{\rm c}$ under both $B=0$ T and $B=12$ T, the SRM at $E_{\rm r1}$ becomes much weaker under $B=12$ T [Fig.~2(b)] while the mode at $E_{\rm r2}$ is hardly affected [Fig.~2(d)]. Similar behavior is also seen at the AF zone boundary along $c$ axis at ${\bf Q}=(1,0,1)$ [Fig.~2(c)], where it is possible to cover the full energy range of the SRM at $E_{\rm r2}$. Since the SRM at $E_{\rm r2}$ is $L$-independent \cite{CZhang2013_PRL}, the modest suppression seen in the energy range $5\leq E\leq10$ meV at ${\bf Q}=(1,0,1)$ also applies to ${\bf Q}=(1,0,0.5)$. On the other hand, the SRM at $E_{\rm r1}$ displays significant $L$-dependence and is much weaker at ${\bf Q}=(1,0,1)$ \cite{CZhang2013_PRL}, nevertheless it is also strongly suppressed  at $B=12$ T, similar to the behavior at ${\bf Q}=(1,0,0.5)$ in Fig.~2(a).

\begin{figure}[t]
	\includegraphics[scale=.45]{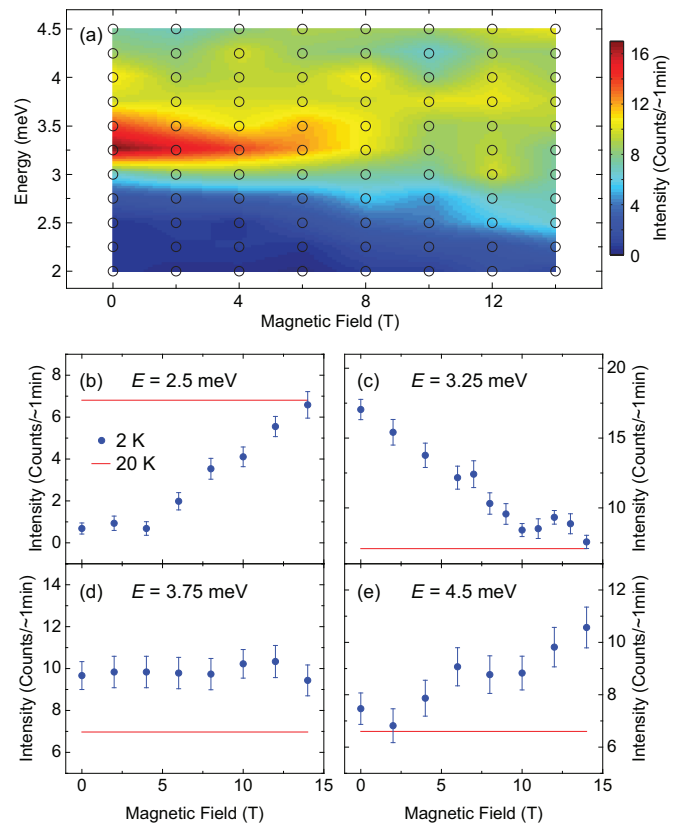}
	\caption{
		(Color online) (a) Color-coded and interpolated magnetic field dependence of low-energy magnetic fluctuations at ${\bf Q}=(1,0,0.5)$ and $T=2$ K, with background subtracted. The circles correspond to points where measurements were taken. Magnetic field dependence of magnetic fluctuations at ${\bf Q}=(1,0,0.5)$ and $T=2$ K for (b) $E=2.5$ meV, (c) $E=3.25$ meV, (d) $E=3.75$ meV and (e) $E=4.5$ meV. The flat red lines are from fit at 20 K, which does not show magnetic field dependence. See Appendix for details on background subtraction.
	}
\end{figure}

Having shown that the SRM at $E_{\rm r2}$ is only modestly suppressed at $B=12$ T, similar to the behavior of SRMs under applied magnetic field in other unconventional superconductors \cite{PBourges1997,Dai2000,JSWen2010_PRB,JZhao2010_PRB,SLi2011_PRB}, we focus on the SRM at $E_{\rm r1}$ which responds much more dramatically to applied field, and study its evolution as a function of applied field. We mapped out the intensity of $T=2$ K magnetic excitations at ${\bf Q}=(1,0,0.5)$ as a function of energy ($2$ meV $\leq E\leq4.5$ meV) and field ($0$ T $\leq B\leq14$ T), as shown in Fig.~3(a). Strong suppression of the SRM at $E_{\rm r1}$ with applied field is immediately apparent. Notably despite the strong suppression in intensity, we do not observe softening for energy of the mode up to $B\approx8$ T, and at higher fields the mode is no longer discernible.

To see how the applied magnetic field affects the SRM at different energies, we show detailed scans from Fig.~3(a) at representative energies in Figs.~3(b)-3(e), compared to the 20 K response (horizontal lines, since the 20 K response is field-independent). At $E=2.5$ meV [Fig.~3(b)], which is inside a superconductivity-induced spin gap at zero-field, magnetic intensity gradually increases with increasing field. This indicates the superconductivity-induced spin gap becomes smaller with applied field [Fig.~3(a)], and is similar to previous observations in optimal-doped BaFe$_{1.9}$Ni$_{0.1}$As$_2$ \cite{JZhao2010_PRB}. At $E=E_{\rm r1}=3.25$ meV [Fig.~3(c)], intensity is quickly suppressed with applied field and plateaus for $B\gtrsim8$ T, despite superconductivity persisting to $B\gtrsim40$ T \cite{SGhannadzadeh2014_PRB}. This behavior is completely different from what was previously seen in BaFe$_{1.9}$Ni$_{0.1}$As$_2$ \cite{JZhao2010_PRB}, where suppression of the SRM tracks suppression of superconductivity under applied magnetic field \cite{JZhao2010_PRB}. At $E=3.75$ meV [Fig.~3(d)], corresponding to a shoulder of the SRM at $E_{\rm r1}$, while the intensity shows clear enhancement relative to the normal state, no significant field dependence is observed. At $E=4.5$ meV [Fig.~3(e)], corresponding to the valley between the two SRMs, the intensity gradually increases with applied field, confirming the valley between $E_{\rm r1}$ and $E_{\rm r2}$ disappears with increasing field [see also Fig.~2(a)].

\begin{figure}[t]
\includegraphics[scale=.45]{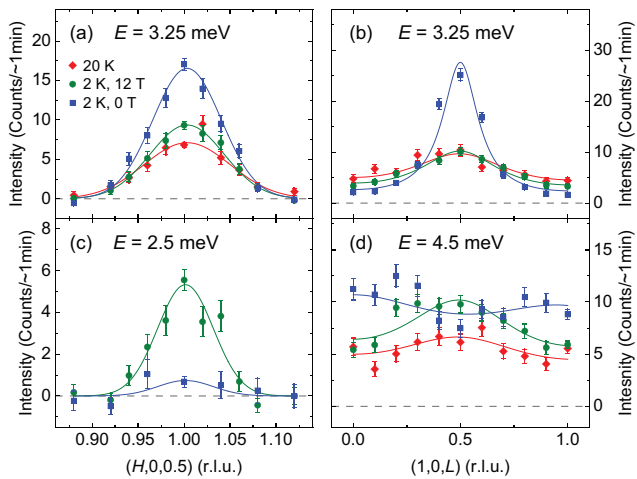}
\caption{
(Color online) (a) Background-subtracted constant-energy scans along $(H,0,0.5)$ for $E=3.25$ meV, (b) along $(1,0,L)$ for $E=3.25$ meV, (c) along $(H,0,0.5)$ for $E=2.5$ meV and (d) along $(1,0,L)$ for $E=4.5$ meV. Solid lines in (a) and (c) are fits to Gaussian peaks, and solid lines in (b) and (d) are fits to lattice Lorentzian peaks. See Appendix for details on background subtraction.
 }
\end{figure}

Further insight into how the low-energy spin dynamics evolve under applied magnetic field can be gained by examining constant-energy scans shown in Fig.~4. At $E=E_{\rm r1}=3.25$ meV, scans along $(H,0,0.5)$ [Fig.~4(a)] and $(1,0,L)$ [Fig.~4(b)] both confirm the strong suppression of the SRM at $E=E_{\rm r1}$ under $B=12$ T. Moreover, correlation length along $L$ for the response at $B=12$ T is significantly shorter compared to $B=0$ T, suggesting the intense SRM at zero-field involving spins in many Fe-As layers is fully suppressed, replaced by fluctuations of the spins that display weak correlations between Fe-As planes. Suppression of the superconductivity-induced spin gap can be clearly seen in Fig.~4(c) at $E=2.5$ meV, while at zero-field there is almost no magnetic signal, a clear peak is observed under $B=12$ T. At $E=4.5$ meV [Fig.~4(d)], which corresponds to $E_{\rm r1}$ at integer $L$-values, and which at $L=0.5$ corresponds to the valley between $E_{\rm r1}$ and $E_{\rm r2}$, display dramatically different $L$-dependence between $B=0$ T and $B=12$ T. At zero-field, we find the magnetic fluctuations to be peaked at integer $L$-values, consistent with previous report \cite{CZhang2013_PRL}. However, under $B=12$ T the magnetic fluctuations peak at $L=0.5$, similar to the normal state response. This dramatic change in $L$-dependence also evidence the SRM at $E_{\rm r1}$, which disperse along $L$ from $E_{\rm r1}=3.25$ meV at $L=0.5$ to $E_{\rm r1}=4.5$ meV at $L=1$, is fully suppressed under $B=12$ T, replaced by magnetic fluctuations that are always centered at $L=0.5$.

\begin{figure}[t]
\includegraphics[scale=.45]{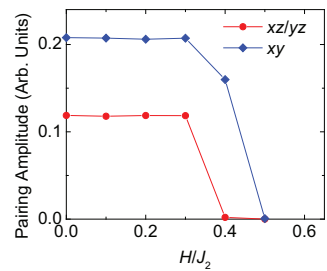}
\caption{
(Color online) Orbital-selective destruction of the superconducting pairing amplitudes by the applied magnetic field in a five-orbital $t$-$J$ model. Shown are the leading pairing channels with $s\pm$ symmetry in $xz/yz$ and $xy$ orbitals. The horizontal axis is the ratio of applied magnetic field $H$ and the next-nearest-neighbor exchange coupling $J_2$. 
 }
\end{figure}

\section{Discussion}

Our results show that the low-energy SRM is strongly suppressed by a magnetic field well below $B_{\rm c2}$, while the high-energy mode is only weakly suppressed. The complete suppression of the low-energy SRM when static magnetic order is gradually enhanced with field [Fig.~1(a)] suggests that it is not directly associated with AF order. Given $T_{\rm c}$ is only weakly modified by a magnetic field much smaller than $B_{\rm c2}$, the suppression of the low-energy SRM is also unlikely to be due to reduction of the dominant superconducting gaps that determines $T_{\rm c} $.

On the other hand, the low-energy SRM is spin-anisotropic while the high-energy one is spin-isotropic~\cite{Zhang_PRB90_140502}. Such a spin-space anisotropy reflects different orbital characters associated with the two resonances, after taking into account the effect of spin-orbit coupling. In fact, theoretical calculation~\cite{Yu_PRB89_024509} has found that the high-energy SRM is mainly associated with the $d_{xy}$ orbital, and the  low-energy one involves $d_{xz}$ and $d_{yz}$ orbitals.
The orbital character of the Fermi surface in NaFe$_{1-x}$Co$_x$As [Fig.~1(a)] \cite{YZhang_PRB2012,MYi_NJP2012} and the anisotropic superconducting gaps [Fig.~1(b)] \cite{QQGe_PRX2013} suggest that the $d_{xy}$ orbital exhibits stronger superconducting pairing whereas $d_{xz}$/$d_{yz}$ has weaker pairing strength, in contrast to FeSe with pairing mainly due to $d_{xz}$/$d_{yz}$ orbitals \cite{POSprau}. Within a five-orbital $t$-$J$ model~\cite{Yu_PRB89_024509}, we studied how the superconducting pairing evolves under a magnetic field. Our main result is summarized in Fig.~5. The pairing strengths of the leading $s\pm$ pairing channels in both the $xz/yz$ and $xy$ orbitals are stable against weak fields, but are reduced when the field becomes strong. Interestingly, the suppression of the pairing amplitudes undergoes in an orbital-selective way. At an intermediate field, the small superconducting gaps associated with $d_{xz}$/$d_{yz}$ orbitals 
become strongly suppressed by the applied magnetic field, while $T_{\rm c}$ is determined by superconducting gaps associated with $d_{xy}$ orbitals, which remain robust for a similar field. The disparate fate of the two SRMs in underdoped NaFe$_{1-x}$Co$_x$As under applied field 
then results from their orbital-selective nature, with the high-energy mode 
associated with $d_{xy}$ orbitals and maintains its intensity, while the low-energy mode involves $d_{xz}$/$d_{yz}$ orbitals and is strongly suppressed.
Finally, we note that suppression of the low-energy SRM by a field well below $B_{\rm c2}$ in underdoped NaFe$_{1-x}$Co$_x$As is reminiscent of amplitude Higgs mode's behavior under magnetic field in superconducting $2H$-NbSe$_2$ \cite{RSooryakumar_PRL1980,RSooryakumar_PRB1981,PBLittlewood_PRB1982}, which also display strong suppression by a magnetic field well below $B_{\rm c2}$ while exhibiting little or no softening of energy of the mode. The field-sensitivity of the Higgs mode in $2H$-NbSe$_2$ is suggested to arise from suppression of the superconducting volume due to the formation of vortices \cite{PBLittlewood_PRB1982}, which cannot account for what we observe in underdoped NaFe$_{1-x}$Co$_x$As. This is because intensity of the SRM at $E_{\rm r2}$, which is reflective of the superconducting volume, is only weakly affected by the magnetic field.

\section{Acknowledgements}
We thank Manh Duc Le and Diana Luc\'{i}a Quintero-Castro for assistance in preliminary neutron scattering measurements, and Ming Yi and Dung-Hai Lee for helpful discussions. The neutron-scattering work at Rice University is supported by the United States DOE, BES under Contract No. DESC0012311 (P.D.). A part of the materials work at Rice is supported by the Robert A. Welch Foundation through Grant No. C-1839 (P.D.). We thank HZB for the allocation of neutron radiation beam time.
\section{Appendix}
\subsection{Background subtraction}

Background for constant-${\bf Q}$ scans should be measured at positions with the same $|{\bf Q}|$ but no magnetic signal. From constant-energy scans shown in Fig.~4 and in previous work \cite{CZhang2013_PRL}, magnetic excitations in underdoped NaFe$_{1-x}$Co$_x$As are relatively sharp along $H$ and broad along $L$, we have therefore chosen ${\bf Q}=(0.8,0,L)$ to measure the background. Constant-{\bf Q} scans before background-subtraction are shown in Fig.~6(a) for ${\bf Q}=(1,0.5)$ and in Fig.~6(c) for ${\bf Q}=(1,0,1)$, together with respective background measurements at ${\bf Q}=(0.8,0,0.902)$ and ${\bf Q}=(0.8,0,1.25)$. The background is then fit to an empirical form and the fit values have been subtracted from results presented in Figs.~2 and 3.

Raw data of constant-energy scans were fit with a Gaussian or a lattice Lorentzian peak plus a linear background, the linear background was constrained to be identical for different temperatures and applied fields. The resulting linear background was subtracted from the raw data, with the results shown in Fig.~4.

\subsection{Field-dependence of normal state excitations}
Figs.~6(b) and (d) respectively show constant-${\bf Q}$ scans at $20$ K for ${\bf Q}=(1,0,0.5)$ and ${\bf Q}=(1,0,1)$, under different applied fields. Similar to previous results on BaFe$_{2-x}$Ni$_x$As$_2$ \cite{MWang2011_PRB}, we do not observe significant field-dependence for the normal state excitations, therefore we combined our data measured at 20 K under different fields.

\begin{figure}[t]
	\includegraphics[scale=.45]{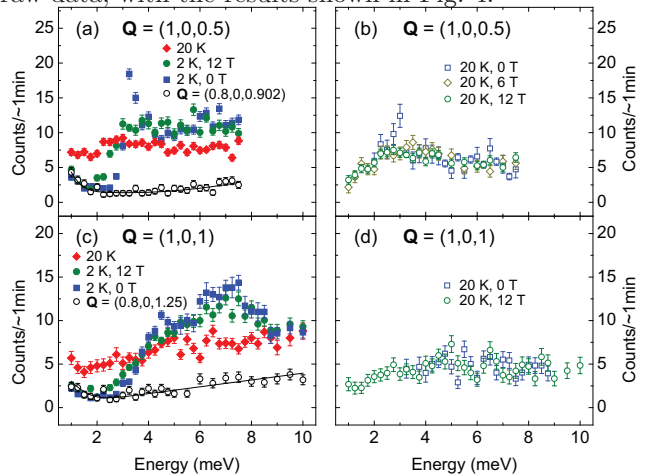}
	\caption{
		(Color online) (a) Constant-${\bf Q}$ scans at ${\bf Q}=(1,0,0.5)$, without background-subtraction. Background measured at ${\bf Q}=(0.8,0,0.902)$ is shown for comparison, and the solid line is an empirical fit to the background. (b) Background-subtracted constant-${\bf Q}$ scans at ${\bf Q}=(1,0,0.5)$ for $T=20$ K under different applied fields. (c) Constant-${\bf Q}$ scans at ${\bf Q}=(1,0,1)$, without background-subtraction. Background measured at ${\bf Q}=(0.8,0,1.25)$ is shown for comparison, and the solid line is an empirical fit to the background. (d) Background-subtracted constant-${\bf Q}$ scans at ${\bf Q}=(1,0,1)$ for $T=20$ K under different applied fields.
	}
\end{figure}


\begin{thebibliography}{}
	
\bibitem{DScalapino2012_RMP} D. J. Scalapino, Rev. Mod. Phys. {\bf 84}, 1383 (2012).	

\bibitem{PDai2015_RMP} Pengcheng Dai, Rev. Mod. Phys. {\bf 87}, 855 (2015).	

\bibitem{Inosov} D. S. Inosov, C. R. Phys. {\bf 17}, 60 (2016).

\bibitem{JRossat-Mignod1991_PC} J. Rossat-Mignod, L. P. Regnault, C. Vettier, P. Bourges, P. Burlet, J. Bossy, J. Y. Henry, and G. Lapertot, Physica C {\bf 185} 86-92 (1991).

\bibitem{CStock2008_PRL} C. Stock, C. Broholm, J. Hudis, H. J. Kang, and C. Petrovic, Phys. Rev. Lett. {\bf 100}, 087001 (2008).

\bibitem{ADChristianson2008_Nature} A. D. Christianson, E. A. Goremychkin, R. Osborn, S. Rosenkranz, M. D. Lumsden, C. D. Malliakas, I. S. Todorov, H. Claus, D. Y. Chung, M. G. Kanatzidis, R. I. Bewley, and T. Guidi, Nature {\bf 456}, 930-932 (2008).

\bibitem{hirschfeld} P. J. Hirschfeld, M. M. Korshunov, and I. I. Mazin, Rep. Prog. Phys. {\bf 74}, 124508 (2011).

\bibitem{MEschrig2006} M. Eschrig, Adv. Phys. {\bf 55}, 47-183 (2006).

\bibitem{YSidis2007} Yvan Sidis, St\'{e}phane Pailh\`{e}s, Vladimir Hinkov, Beno\^{i}t Fauqu\'{e}, Clemens Ulrich, Lucia Capogna, Alexandre Ivanov, Louis-Pierre Regnault, Bernhard Keimer, Philippe Bourges, C. R. Physique {\bf 8}, 745-762 (2007).

\bibitem{Dai2000} Pengcheng Dai, H. A. Mook, G. Aeppli, S. M. Hayden, and F. Do\u{g}an, Nature {\bf 406}, 965-968 (2000).

\bibitem{PBourges1997} P. Bourges, H. Casalta, L. P. Regnault J. Bossy, P. Burlet, C. Vettier, E. Beaugnon, P.Gautier-Picard, and R.Tournier, Physica B {\bf 234-236}, 830-831 (1997).

\bibitem{JSWen2010_PRB} Jinsheng Wen, Guangyong Xu, Zhijun Xu, Zhi Wei Lin, Qiang Li, Ying Chen, Songxue Chi, Genda Gu, and J. M. Tranquada, Phys. Rev. B {\bf 81}, 100513 (2010).

\bibitem{JZhao2010_PRB} Jun Zhao, Louis-Pierre Regnault, Chenglin Zhang, Miaoying Wang, Zhengcai Li, Fang Zhou, Zhongxian Zhao, Chen Fang, Jiangping Hu, and Pengcheng Dai, Phys. Rev. B {\bf 81}, 180505 (2010).

\bibitem{SLi2011_PRB} Shiliang Li, Xingye Lu, Meng Wang, Hui-qian Luo, Miaoyin Wang, Chenglin Zhang, Enrico Faulhaber, Louis-Pierre Regnault, Deepak Singh, and Pengcheng Dai, Phys. Rev. B {\bf 84}, 024518 (2011).

\bibitem{GRStewart} G. R. Stewart, Rev. Mod. Phys. {\bf 83}, 1589 (2011).

\bibitem{MYi_npj_QM} Ming Yi, Yan Zhang, Zhi-Xun Shen, and Donghui Lu, npj Quantum Materials {\bf 2}, 57 (2017).

\bibitem{ZPYin_NM} Z. P. Yin, K. Haule, and G. Kotliar, Nat. Mater. {\bf 10}, 932-935 (2011).

\bibitem{deMedici_PRL} Luca de' Medici, Gianluca Giovannetti, and Massimo Capone, Phys. Rev. Lett. {\bf 112}, 177001 (2014).

\bibitem{MYi_PRL2013} M. Yi, D. H. Lu, R. Yu, S. C. Riggs, J.-H. Chu, B. Lv, Z. K. Liu, M. Lu, Y.-T. Cui, M. Hashimoto, S.-K. Mo, Z. Hussain, C. W. Chu, I. R. Fisher, Q. Si, and Z.-X. Shen, Phys. Rev. Lett. {\bf 110}, 067003 (2013).

\bibitem{RYu_PRL2013} Rong Yu, and Qimiao Si, Phys. Rev. Lett. {\bf 110}, 146402 (2013).

\bibitem{EMNica_npj_QM} Emilian M. Nica, Rong Yu, and Qimiao Si, npj Quantum Materials {\bf 2}, 24 (2017).

\bibitem{POSprau} P. O. Sprau, A. Kostin, A. Kreisel, A. E. B{\"o}hmer, V. Taufour, P. C. Canfield, S. Mukherjee, P. J. Hirschfeld, B. M. Andersen, J. C. S{\'e}amus Davis, Science {\bf 357}, 75-80 (2017).

\bibitem{YLi_PRL2016} Yu Li, Zhiping Yin, Xiancheng Wang, David W. Tam, D. L. Abernathy, A. Podlesnyak, Chenglin Zhang, Meng Wang, Lingyi Xing, Changqing Jin, Kristjan Haule, Gabriel Kotliar, Thomas A. Maier, and Pengcheng Dai, Phys. Rev. Lett. {\bf 116}, 247001 (2016).

\bibitem{CZhang2013_PRL} Chenglin Zhang, Rong Yu, Yixi Su, Yu Song, Miaoyin Wang, Guotai Tan, Takeshi Egami, J. A. Fernandez-Baca, Enrico Faulhaber, Qimiao Si, and Pengcheng Dai, Phys. Rev. Lett. {\bf 111}, 207002 (2013).

\bibitem{PSteffens2013_PRL} P. Steffens, C. H. Lee, N. Qureshi, K. Kihou, A. Iyo, H. Eisaki, and M. Braden, Phys. Rev. Lett. {\bf 110}, 137001 (2013).

\bibitem{CZhang2013_PRB} Chenglin Zhang, Mengshu Liu, Yixi Su, Louis-Pierre Regnault, Meng Wang, Guotai Tan, Th. Br\"{u}ckel, Takeshi Egami, and Pengcheng Dai, Phys. Rev. B {\bf 87}, 081101 (2013).

\bibitem{DHu2017_PRB} Ding Hu, Wenliang Zhang, Yuan Wei, Bertrand Roessli, Markos Skoulatos, Louis Pierre Regnault, Genfu Chen, Yu Song, Huiqian Luo, Shiliang Li, and Pengcheng Dai, Phys. Rev. B {\bf 96}, 180503 (2017).

\bibitem{WLv2014_PRB} Weicheng Lv, Adriana Moreo, and Elbio Dagotto, Phys. Rev. B {\bf 89}, 104510 (2014).

\bibitem{MWang2016_PRB} Meng Wang, M. Yi, H. L. Sun, P. Valdivia, M. G. Kim, Z. J. Xu, T. Berlijn, A. D. Christianson, Songxue Chi, M. Hashimoto, D. H. Lu, X. D. Li, E. Bourret-Courchesne, Pengcheng Dai, D. H. Lee, T. A. Maier, and R. J. Birgeneau, Phys. Rev. B {\bf 93}, 205149 (2016).

\bibitem{SVBorisenko_NP2016} S. V. Borisenko, D. V. Evtushinsky, Z.-H. Liu, I. Morozov, R. Kappenberger, S. Wurmehl, B. B\"{u}chner, A. N. Yaresko, T. K. Kim, M. Hoesch, T. Wolf, and N. D. Zhigadlo, Nat. Phys. {\bf 12} 311-317 (2016).

\bibitem{YSong2016_PRB} Yu Song, Haoran Man, Rui Zhang, Xingye Lu, Chenglin Zhang, Meng Wang, Guotai Tan, L.-P. Regnault, Yixi Su, Jian Kang, Rafael M. Fernandes, and Pengcheng Dai, Phys. Rev. B {\bf 94}, 214516 (2016).

\bibitem{MMKorshunov2013_JSNM} M. M. Korshunov, Y. N. Togushova, I. Eremin, and P. J. Hirschfeld, J. Supercond. Nov. Magn. {\bf 26}, 2873-2874 (2013).

\bibitem{DDScherer2017} Daniel D. Scherer, and Brian M. Andersen, arXiv:1711.02460 (2017).


\bibitem{QQGe_PRX2013} Q. Q. Ge, Z. R. Ye, M. Xu, Y. Zhang, J. Jiang, B. P. Xie, Y. Song, C. L. Zhang, Pengcheng Dai, and D. L. Feng, Phys. Rev. X {\bf 3}, 011020 (2013).

\bibitem{GTan2016_PRB} Guotai Tan, Yu Song, Chenglin Zhang, Lifang Lin, Zhuang Xu, Tingting Hou, Wei Tian, Huibo Cao, Shiliang Li, Shiping Feng, and Pengcheng Dai, Phys. Rev. B {\bf 94}, 014509 (2016).

\bibitem{SVCarr2016_PRB} Scott V. Carr, Chenglin Zhang, Yu Song, Guotai Tan, Yu Li, D. L. Abernathy, M. B. Stone, G. E. Granroth, T. G. Perring, and Pengcheng Dai, Phys. Rev. B {\bf 93}, 214506 (2016).

\bibitem{CZhang2016_PRB} Chenglin Zhang, Weicheng Lv, Guotai Tan, Yu Song, Scott V. Carr, Songxue Chi, M. Matsuda, A. D. Christianson, J. A. Fernandez-Baca, L. W. Harriger, and Pengcheng Dai, Phys. Rev. B {\bf 93}, 174522 (2016).

\bibitem{SLi2009_PRB} Shiliang Li, Clarina de la Cruz, Q. Huang, G. F. Chen, T.-L. Xia, J. L. Luo, N. L. Wang, and Pengcheng Dai, Phys. Rev. B {\bf 80}, 020504 (2009).

\bibitem{CDhital2012_PRL} Chetan Dhital, Z. Yamani, Wei Tian, J. Zeretsky, A. S. Sefat, Ziqiang Wang, R. J. Birgeneau, and Stephen D. Wilson, Phys. Rev. Lett. {\bf 108}, 087001 (2012).

\bibitem{YSong2013_PRB} Yu Song, Scott V. Carr, Xingye Lu, Chenglin Zhang, Zachary C. Sims, N. F. Luttrell, Songxue Chi, Yang Zhao, Jeffrey W. Lynn, and Pengcheng Dai, Phys. Rev. B {\bf 87}, 184511 (2013).

\bibitem{DTam2017_PRB} David W. Tam, Yu Song, Haoran Man, Sky C. Cheung, Zhiping Yin, Xingye Lu, Weiyi Wang, Benjamin A. Frandsen, Lian Liu, Zizhou Gong, Takashi U. Ito, Yipeng Cai, Murray N. Wilson, Shengli Guo, Keisuke Koshiishi, Wei Tian, Bassam Hitti, Alexandre Ivanov, Yang Zhao, Jeffrey W. Lynn, Graeme M. Luke, Tom Berlijn, Thomas A. Maier, Yasutomo J. Uemura, and Pengcheng Dai, Phys. Rev. B {\bf 95}, 060505 (2017).

\bibitem{WWang_PRB2017} Weiyi Wang, J. T. Park, Rong Yu, Yu Li, Yu Song, Zongyuan Zhang, Alexandre Ivanov, Jiri Kulda, and Pengcheng Dai, Phys. Rev. B {\bf 95}, 094519 (2017).

\bibitem{ADChristianson2009_PRL} A. D. Christianson, M. D. Lumsden, S. E. Nagler, G. J. MacDougall, M. A. McGuire, A. S. Sefat, R. Jin, B. C. Sales, and D. Mandrus, Phys. Rev. Lett. {\bf 103}, 087002 (2009).

\bibitem{JLarsen2015_PRB} J. Larsen, B. Mencia Uranga, G. Stieper, S. L. Holm, C. Bernhard, T. Wolf, K. Lefmann, B. M. Andersen, and C. Niedermayer, Phys. Rev. B {\bf 91}, 024504 (2015).

\bibitem{SGhannadzadeh2014_PRB} S. Ghannadzadeh, J. D. Wright, F. R. Foronda, S. J. Blundell, S. J. Clarke, and P. A. Goddard, Phys. Rev. B {\bf 89}, 054502 (2014).

\bibitem{MWang2011_PRB} Miaoyin Wang, Huiqian Luo, Meng Wang, Songxue Chi, Jose A. Rodriguez-Rivera, Deepak Singh, Sung Chang, Jeffrey W. Lynn, and Pengcheng Dai, Phys. Rev. B {\bf 83}, 094516 (2011).

\bibitem{Zhang_PRB90_140502} Chenglin Zhang, Yu Song, L.-P. Regnault, Yixi Su, M. Enderle, J. Kulda, Guotai Tan, Zachary C. Sims, Takeshi Egami, Qimiao Si, and Pengcheng Dai, Phys. Rev. B {\bf 90}, 140502 (2014).

\bibitem{Yu_PRB89_024509} Rong Yu, Jian-Xin Zhu, and Qimiao Si, Phys. Rev. B {\bf 89}, 024509 (2014).

\bibitem{YZhang_PRB2012} Y. Zhang, C. He, Z. R. Ye, J. Jiang, F. Chen, M. Xu, Q. Q. Ge, B. P. Xie, J. Wei, M. Aeschlimann, X. Y. Cui, M. Shi, J. P. Hu, and D. L. Feng, Phys. Rev. B {\bf 85}, 085121 (2012).

\bibitem{MYi_NJP2012} M. Yi, D. H. Lu, R. G. Moore, K. Kihou, C.-H. Lee, A. Iyo, H. Eisaki, T. Yoshida, A. Fujimori, and Z.-X. Shen, New J. Phys. {\bf 14}, 073019 (2012).




\bibitem{RSooryakumar_PRL1980} R. Sooryakumar, and M. V. Klein, Phys. Rev. Lett. {\bf 45}, 660 (1980).
	
\bibitem{RSooryakumar_PRB1981} R. Sooryakumar, and M. V. Klein, Phys. Rev. B {\bf 23}, 3213 (1981).


\bibitem{PBLittlewood_PRB1982} P. B. Littlewood, and C. M. Varma, Phys. Rev. B {\bf 26}, 4883 (1982).


\end{thebibliography}
\end{document}